%% file: root.tex
\documentclass[a4paper, 10pt, conference]{ieeeconf}

\newif\ifoagmfinalcopy

\oagmfinalcopytrue  

\IEEEoverridecommandlockouts                       

\usepackage{graphicx} 
\usepackage{epsfig} 
\usepackage{mathptmx} 
\usepackage{times} 
\usepackage{amsmath} 
\usepackage{amssymb}  
\usepackage{lineno}
\usepackage{anyfontsize}
\usepackage{tikzpagenodes}
\usepackage{background}
\usepackage{multirow}
\usepackage{microtype}
\usepackage{csquotes}
\usepackage{mathtools}
\usepackage{siunitx}
\usepackage{booktabs}
\usepackage{pgfplots}
\pgfplotsset{compat=1.17}
\usepackage[ruled,linesnumbered]{algorithm2e}
\usepackage{tikz}
\usetikzlibrary{spy}
\usepackage{glossaries}
\usepackage{ifthen}
\usetikzlibrary{shapes.geometric}
\usepackage{xcolor}
\usetikzlibrary{shapes.arrows}
\definecolor{photoncolor}{rgb}{0.65, 0.16, 0.16}
\definecolor{light}{HTML}{bdbdbd}
\definecolor{normalanno}{HTML}{446699}
\definecolor{threebythree}{HTML}{543639}
\newcommand{\networkblock}[3]{%
	\pgfmathsetmacro{\cubex}{#1}
	\pgfmathsetmacro{\cubey}{#2}
	\pgfmathsetmacro{\cubez}{#3}
	\draw[line join=bevel, ultra thick,draw=photoncolor,fill=light] (0,0,0) -- ++(-\cubex,0,0) -- ++(0,-\cubey,0) -- ++(\cubex,0,0) -- cycle;
	\draw[line join=bevel, ultra thick,draw=photoncolor,fill=light] (0,0,0) -- ++(0,0,-\cubez) -- ++(0,-\cubey,0) -- ++(0,0,\cubez) -- cycle;
	\draw[line join=bevel, ultra thick,draw=photoncolor,fill=light] (0,0,0) -- ++(-\cubex,0,0) -- ++(0,0,-\cubez) -- ++(\cubex,0,0) -- cycle;
}
\input{acronyms}
\tikzdeclarecoordinatesystem{image}{%
  \tikzset{image cs/.cd, #1}%
    \pgfpointdiff%
      {\pgfpointanchor{\graphicname}{south west}}%
      {\pgfpointanchor{\graphicname}{north east}}%
    \pgfgetlastxy\graphicwidth\graphicheight%
    \pgfmathparse{\graphicx}%
    \ifpgfmathunitsdeclared\def\graphicwidth{1}\fi%
    \pgfmathparse{\graphicy}%
    \ifpgfmathunitsdeclared\def\graphicheight{1}\fi%
    \pgfpointadd{\pgfpointanchor{\graphicname}{south west}}%
      {\pgfpoint{(\graphicx)*\graphicwidth}{(\graphicy)*\graphicheight}}%
}
\tikzset{image cs/.cd,
  x/.store in=\graphicx, y/.store in=\graphicy,
  image/.store in=\graphicname
}
\ifoagmfinalcopy
\backgroundsetup{color=white}
\else

\linenumbers

\SetBgContents{}
\SetBgPosition{current page.north west}
\SetBgOpacity{0.5}
\SetBgAngle{0.0}
\SetBgScale{1.0}

\fi
\DeclareMathOperator*{\argmin}{arg\,min}

\newcommand{\R}{\mathbb{R}}
\newcommand{\optimal}[1]{#1^{*}}
\newcommand{\norm}[1]{\left\lVert#1\right\rVert}

\makeatletter
\let\NAT@parse\undefined
\makeatother
\usepackage[colorlinks=true,citecolor=photoncolor,linkcolor=normalanno]{hyperref}

\title{\LARGE \bf Computed Tomography Reconstruction\\Using Generative Energy-Based Priors}

\ifoagmfinalcopy
\author{Martin Zach\textsuperscript{1}, Erich Kobler\textsuperscript{2}, and Thomas Pock\textsuperscript{1}
\thanks{%
    \textsuperscript{1}Institute of Computer Graphics and Vision, Graz University of Technology, 8010 Graz, %
    Austria \texttt{\small \{martin.zach,pock\}@icg.tugraz.at}%
}%
\thanks{%
    \textsuperscript{2}Institute of Computer Graphics, Johannes Kepler University Linz, 4040 Linz, Austria {\tt\small erich.kobler@jku.at}}%
}
\else
\author{Anon, Ymous}
\fi
\begin{document}
\maketitle
\begin{abstract}
In the past decades, \gls{ct} has established itself as one of the most important imaging techniques in medicine.
Today, the applicability of \gls{ct} is only limited by the deposited radiation dose, reduction of which manifests in noisy or incomplete measurements.
Thus, the need for robust reconstruction algorithms arises.
In this work, we learn a parametric regularizer with a global receptive field by maximizing it's likelihood on reference CT data.
Due to this unsupervised learning strategy, our trained regularizer truly represents higher-level domain statistics, which we empirically demonstrate by synthesizing \gls{ct} images.
Moreover, this regularizer can easily be applied to different \gls{ct} reconstruction problems by embedding it in a variational framework, which increases flexibility and interpretability compared to feed-forward learning-based approaches. 
In addition, the accompanying probabilistic perspective enables experts to explore the full posterior distribution and may quantify uncertainty of the reconstruction approach.
We apply the regularizer to limited-angle and few-view \gls{ct} reconstruction problems, where it outperforms traditional reconstruction algorithms by a large margin. 
\end{abstract}
\glsresetall
\section{INTRODUCTION}
Throughout the past decades, \gls{ct} has become an invaluable tool in diagnostic radiology.
However, along with its ever-increasing usage have come concerns about the associated risks from ionizing radiation exposure~\cite{brenner_computed_2007}.
Approaches that try to remedy this problem include hardware measures such as tube current reduction or modulation (for instance in the form of automatic exposure control~\cite{Sderberg2010}), adaptive section collimation~\cite{Deak2009}, or angular under-sampling~\cite{Chen2008,chen_sparsect_2019}.
Such measures are now standard in clinical \gls{ct} systems, but require robust reconstruction algorithms.

Classical \gls{ct} reconstruction algorithms include \gls{fbp}~\cite{buzug_computed_2008,feldkamp_practical_1984}, which has been superseded by more robust iterative algebraic reconstruction techniques~\cite{saad_iterative_2000,wang_vannier_cheng_1999} in clinical practice.
In light of dose reduction, these algorithms may be equipped with prior knowledge to increase reconstruction quality of low-dose scans.
Traditional, hand-crafted regularizers, such as \gls{tv}~\cite{rudin_nonlinear_1992} and extensions such as \gls{tgv}~\cite{bredies_total_2010}, typically encode regularity assumptions of the reconstruction, such as sparsity of gradients.
These hand-crafted regularizers have been used extensively and successfully in reconstruction problems~\cite{chen_limited_2013,liu_total_2014,zhang_few-view_2013}, however they do not fully model the a-priori available information.
To capture also higher-order image statistics, the idea of learning a regularizer from data emerged~\cite{zhu_filters_1998,roth_fields_2005,kobler_total_2020}.
Although these learning-based approaches are now dominant in many fields, such models have classically focused on modeling local statistics and leave much to be desired in modeling global dependencies.

From a statistical point of view, any regularizer \( R \) induces a Gibbs-Boltzmann distribution
\begin{equation}
    p_R(x) = \frac{\exp(-R(x))}{\int_{\mathcal{X}}\exp(-R(\xi))\ \mathrm{d}\xi},%
    \label{eq:gibbs}
\end{equation}
where \( \mathcal{X} \) is the space of all possible images.
Ideally, samples \( x \sim p_R \) should be indistinguishable from samples from the underlying reference distribution, which is hardly possible for hand-crafted regularizers.
\begin{figure}
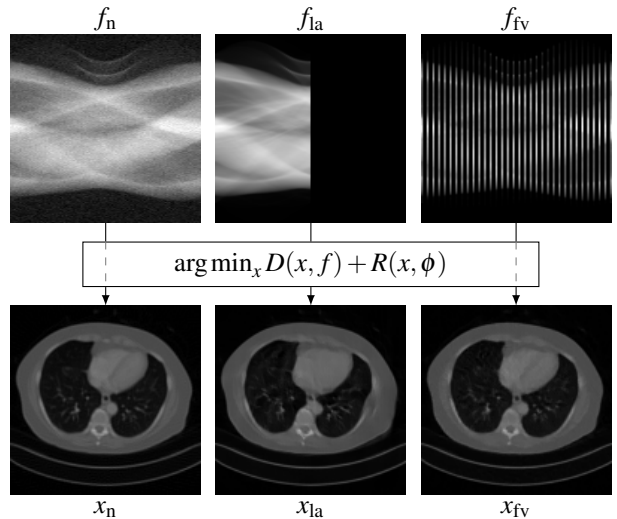

    \centering
    \begin{tikzpicture}
        \def\imwidth{2.5cm}
        \def\pad{0.2cm}
        \node (anno) [rectangle, minimum width=6cm, draw] at (2 * \imwidth + 2 * \pad, -1.8) {\( \argmin_{x} D(x, f) + R(x, \phi) \)};
        \foreach [count=\i] \imname/\anno in {noisy/n,  limited-angle/la, undersampled/fv} {
            \pgfmathsetlengthmacro{\xx}{\i * (\imwidth + \pad)}
            \node at (\xx, 1.45) {\( f_{\text{\anno}} \)};
            \node [outer sep=0, inner sep=0] (f\anno) at (\xx, 0) {\includegraphics[width=\imwidth, height=\imwidth]{./figures/header/sinograms/\imname.png}};
            \node [outer sep=0, inner sep=0] (x\anno) at (\xx, -3.6) {\includegraphics[width=\imwidth, height=\imwidth]{./figures/header/reconstructions/\imname.png}};
            \node at (\xx, -5.05) {\( x_{\text{\anno}} \)};
            \draw (f\anno.south) -- (f\anno.south |- anno.north);
            \draw [-latex] (f\anno.south |- anno.south) -- (x\anno.north);
            \ifthenelse{\i=2}{}{\draw [gray, dashed] (f\anno.south |- anno.north) -- (f\anno.south |- anno.south);}
        }
    \end{tikzpicture}
    \caption{%
        Our proposed method is able to reconstruct images from noisy, limited-angle and few-view measurements (denoted by the subscripts \( \text{n}, \text{la}, \text{fv} \)) satisfactorily.
    }
    \label{fig:header}
\end{figure}

In this work, we propose a novel generatively trained regularizer utilizing a global receptive field that yields high-quality reconstructions even in case of strong noise or heavily undersampled measurements.
In Fig.~\ref{fig:header}, we show how our model is able to satisfactorily reconstruct \gls{ct} images from noisy (i.e.\ low tube current) and incomplete (i.e.\ limited-angle or few-view) data without observable artifacts.
In fact, using this regularizer we can synthesize naturally appearing \gls{ct} images \emph{without any} data (see Fig.~\ref{fig:map}).
In contrast to feed-forward formulations~\cite{rushil_lose_2018,chen_low-dose_2017}, we cast the reconstruction as a variational problem.
This helps interpretability of the trained regularizer by means of analyzing its induced distribution as well as the posterior distribution of any type of reconstruction problem.
We apply a trained model to limited-angle and few-view reconstruction problems, and compare our approach quantitatively and qualitatively with traditional reconstruction algorithms.
In addition, we perform experiments which leverage the probabilistic nature of our approach, such as prior and posterior sampling.

To summarize, we
\begin{itemize}
    \item define a novel network architecture capable of synthesizing natural \gls{ct} images without measurement data,
    \item demonstrate that our regularizer outperforms classical algorithms in typical reconstruction problems, and
    \item show that our probabilistic approach allows to compute the pixel-wise posterior-variance, which in turn is related to uncertainty quantification.
\end{itemize}

\section{RELATED WORK}
\subsection{Learning-based CT Reconstruction}
In recent years, there has been a strong shift from hand-crafted regularizers towards data-driven reconstruction schemes.
The learning-based methods can be applied in the sinogram domain~\cite{rushil_lose_2018,ghani_cnn_2018}, such that the final image can be reconstructed using traditional reconstruction algorithms.
Alternatively, a preliminary reconstruction may be computed using the (noisy and possibly incomplete) sinogram, which can subsequently be enhanced by a trained \gls{cnn}~\cite{chen_low-dose_2017}.
An alternative learning-based reconstruction approach is to learn a direct mapping from the data domain to the image domain~\cite{zhu_image_2018}.
However, this requires to learn a wealth of parameters solely to compute an approximate inverse of the forward acquisition operator.
Another recently popularized approach is to learn an unrolled iterative reconstruction algorithm~\cite{hammernik_learning_2017,adler_learned_2018,kobler_sparse_2020}.
Whilst the results look promising, we point out that such approaches typically assume a particular acquisition setup and, at inference time, can only be applied in settings that are very similar to the training setting.
\subsection{Generative Models as Regularizers in Medical Imaging}
\glspl{ebm} have a long history in the field of image processing~\cite{lecun_tutorial_2006}.
However, only recently some works~\cite{du_implicit_2019,nijkamp_anatomy_2019} have explored their generative capabilities, rivaling the performance of \glspl{gan}.
While \glspl{gan} have been used as an implicit prior for reconstruction problems in medical imaging (e.g.\ \cite{adler_deep_2018}), to the best of our knowledge, using \glspl{ebm} capable of synthesizing natural images at full-scale as regularizers in medical imaging is still largely unexplored.

\section{METHODOLOGY}
In this work, we represent \gls{ct} images of size \( n = n_w \times n_h \) pixels as vectors \( x \in \R^n \).
The subsequent analysis easily generalizes to image data in any dimensions.
Acquiring \( n_\theta \) projections with \( n_d \) detector elements, the post-log sinogram \( f \in \R^m \) of size \( m = n_\theta \times n_d \) is given by
\begin{equation}
    f = Ax + \eta,%
    \label{eq:lip}
\end{equation}
where \( A \colon \R^n \to \R^m \) is the acquisition operator, and \( \eta \in \R^m \) represents the additive measurement noise, summarizing photon statistics, thermal noise in the measurement channels, and pre-processing steps.
The linear acquisition operator \( A \) is defined by the geometry of the measurement setup, and throughout this work we assume that both \( A \) and \( \eta \) can be characterized up to reasonable precision.
\subsection{Bayesian Modeling}
To account for measurement uncertainties and missing data in the observations \( f \), we adopt a rigorous statistical interpretation of~\eqref{eq:lip}.
Bayes' Theorem relates the posterior probability \( p(x \mid f) \) to the data-likelihood \( p(f \mid x) \) and the prior \( p(x) \) by
\begin{equation}
    p(x \mid f) \propto p(f \mid x) p(x).
    \label{eq:bayes}
\end{equation}
Here, \( p(x \mid f) \) quantifies the belief in a solution \( x \) given a datum \( f \).
In the negative log-domain,~\eqref{eq:bayes} is transformed to
\begin{equation}
    E(x, f) \coloneqq D(x, f) + R(x),%
    \label{eq:variational problem}
\end{equation}
where we identify the \emph{data-fidelity} term \( D \colon \R^n \times \R^m \to \R^+ \) modeling the negative log-likelihood \( -\log p(f \mid x) \), and the \emph{regularizer} \( R \colon \R^n \to \R \) modeling the negative log-prior \( -\log p(x) \).
The \emph{energy} \( E \colon \R^n \times \R^m \to \R \) assigns a scalar \( E(x, f) \) to any \( (x, f) \)-pair, and in the sense of~\eqref{eq:bayes} is interpreted as the negative log-posterior \( -\log p(x \mid f) \).

Typically, \( D \) makes use of the forward operator \( A \) to quantify the agreement between the reconstruction of \( x \) and the measured data \( f \).
\( R \) may for instance represent the \gls{tv} semi-norm~\cite{rudin_nonlinear_1992}, which is well known to favor piece-wise constant solutions.
For the sake of simplicity, we assume \( \eta \) to be Gaussian, and consequently set \( D(x, f) = \frac{1}{2\sigma^2} \norm{Ax - f}^2 \), where \( \sigma^2 \) denotes the variance of \( \eta \).
We discuss the choice of \( R \) in the next section.
\subsection{Parameter Identification}
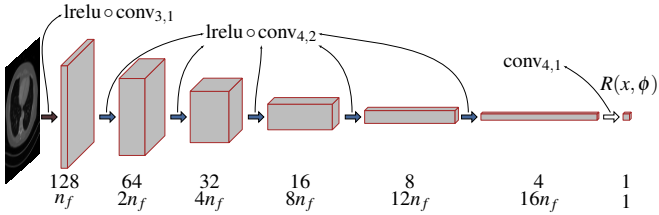
\begin{figure}\centering\scalebox{0.17}{
\begin{tikzpicture}[font=\fontsize{45}{54}\selectfont]
	\foreach [count=\i] \offset/\xsize/\ysize/\annores/\annofeat in {
		1/0.5/8/128/n_f,
		7/2/6/64/2n_f,
		13.5/3/4/32/4n_f,
		21.5/5/2/16/8n_f,
		31/7/1/8/12n_f,
		42/9/0.5/4/16n_f,
		44.5/0.5/0.5/1/1
	}
	{
		\begin{scope}[shift={(\offset, \ysize/2)}]
			\networkblock{\xsize}{\ysize}{\ysize / 1.5}
			\pgfmathsetmacro{\xx}{cos(30) * \ysize / 1.5 / 2}
			\pgfmathsetmacro{\yy}{sin(30) * \ysize / 1.5 / 2}
			\node at (-\xsize/2, -5-\ysize/2) {\( \annores \)};
			\node at (-\xsize/2, -6.6-\ysize/2) {\( \annofeat \)};
		\end{scope}
	}
	\foreach [count=\ii] \offset in {-1,3.5,9,15,22.5,31.5,42.5}
	{
		\ifthenelse{\ii=1}{
				\node (a\ii) [
					draw, fill=threebythree, single arrow,
					minimum height=12mm,
					minimum width=8mm,
					single arrow head extend=2mm, anchor=west, outer sep=2mm,
				] at (\offset - .2, 0){};
			}{
				\ifthenelse{\ii<7}{
					\node (a\ii) [
						draw, fill=normalanno, single arrow,
						minimum height=12mm,
						minimum width=8mm,
						single arrow head extend=2mm, anchor=west, outer sep=2mm,
					] at (\offset -.2, 0){};
				}{
					\node (a\ii) [
						draw, single arrow,
						minimum height=12mm,
						minimum width=8mm,
						single arrow head extend=2mm, anchor=west, outer sep=2mm,
					] at (\offset -.2, 0){};
				}
			}
	}
 	\begin{scope}[overlay]
	\node (a1 anno) at (5, 8) {\( \operatorname{lrelu} \circ \operatorname{conv}_{3, 1} \)};
	\node (a3 anno) at (16, 6.5) {\( \operatorname{lrelu} \circ \operatorname{conv}_{4, 2} \)};
	\foreach \i/\anchor/\inangle in {2/west/190,3/south west/220,4/south/-90,5/south east/-30,6/east/-10}
	{
		\draw [ultra thick, -latex] (a\i.north) to [out=90,in=\inangle] (a3 anno.\anchor);
	}
	\node (a7 anno) at (37, 4) {\( \operatorname{conv}_{4, 1} \)};
	\draw [ultra thick, -latex] (a1.north) to [out=90, in=210] (a1 anno.west);
	\draw [ultra thick, -latex] (a7.north) to [out=120, in=-30] (a7 anno.east);
	\end{scope}
	\node at (-2.5, 0.5) [yslant=0.9, xscale=0.3] {\includegraphics[width=9cm]{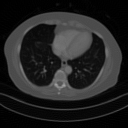}};
	\node at (44.5, 2.5)  {\( R(x, \phi) \)};
\end{tikzpicture}}
\caption{%
    Our proposed architecture follows a typical encoder structure.
    The subscripts specify filter size and stride and the annotations show the spatial resolution of the feature maps and the number of features.
}%
\label{fig:network}
\end{figure}

Although many hand-crafted choices for \( R \) exist, such as \gls{tgv}~\cite{bredies_total_2010} or wavelet-based approaches~\cite{Dong2012}, it is generally agreed upon that modeling higher order image statistics should be based on learning~\cite{zhu_filters_1998}.
In contrast to the widely adopted feed-forward approaches, in this work we retain the variational structure to allow statistical interpretation.
To account for the parameters, we extend~\eqref{eq:variational problem} to
\begin{equation}
    E(x, f, \phi) \coloneqq D(x, f) + R(x, \phi),%
    \label{eq:variational problem parametrized}
\end{equation}
where \( R \colon \R^n \times \Phi \to \R \) is parametrized by \( \phi \) in the set of feasible parameters \( \Phi \).
We illustrate our particular choice of \( R \) (for two-dimensional input images) in Fig.~\ref{fig:network} and emphasize that the input image is reduced to a scalar only by means of (strided) convolutions.
Here, \( \phi \) summarizes the convolution kernels and biases, and \( \Phi \) reduces to \( \R^{n_p} \), where \( n_p \) is the total number of parameters.

The Bayesian separation of of data-likelihood and prior allows us to train our regularizer \emph{generatively} without any measurement data as follows.
We denote by \( p_\phi \) the Gibbs-Boltzmann distribution of \( R(\cdot, \phi) \) in the sense of~\eqref{eq:gibbs}, to emphasize the dependence on the parameters.
Assuming access to a distribution \( p_x \) of reference \gls{ct} images, we identify the optimal parameters \( \optimal{\phi} \) by minimize the negative log-likelihood
\begin{equation}
     \optimal{\phi} \in \argmin_{\phi \in \Phi} \{ \Gamma(\phi) \coloneqq \mathbb{E}_{x\sim p_x}[-\log p_\phi(x)] \}.%
     \label{eq:nll}
\end{equation}
In the seminal work of~\cite{hinton_training_2002}, it is shown that the gradient of \eqref{eq:nll} with respect to the parameters \( \phi \) is given by
\begin{equation}
    \nabla_1 \Gamma(\phi) = \mathbb{E}_{x^+ \sim p_x}[\nabla_2 R(x^+, \phi)] - \mathbb{E}_{x^-\sim p_\phi}[\nabla_2 R(x^-, \phi)],%
    \label{eq:grad nll}
\end{equation}
where \( \nabla_l \) denotes the gradient w.r.t.\ the \( l \)-th argument.
We discuss the estimation of the expectations in both terms extensively in Sec.~\ref{sec:sampling}.

We highlight that~\eqref{eq:nll} does not require any \( (x, f) \)-pairs.
That is, for training we do not require access to measurement data but only to (the usually much more ubiquitous) reference images.
Moreover, a trained regularization model serves as a drop-in replacement for hand-crafted regularizers for any reconstruction problem by adapting the data-fidelity~\( D \) to account for a particular forward operator~\( A \) and noise statistics.
\subsection{Model Sampling} \label{sec:sampling}
While the first term in~\eqref{eq:grad nll} is easily approximated given any dataset, the second term requires sampling the induced model distribution, which is known to be hard in high dimensions~\cite{brooks_handbook_2011}.
For any reasonably sized image \( x \in \R^n \) computing the partition function is infeasible, hence the distribution has to be approximated using \gls{mcmc} techniques.
In this work, we utilize the \gls{ula}~\cite{roberts_exponential_1996, roberts_optimal_1998, rossky_brownian_1978}, which makes use of the gradient of the underlying probability density function to improve mixing times of the Markov chains.
The \gls{ula} algorithm read as
\begin{equation}
    x^k \sim \mathcal{N}(x^{k - 1} + \frac{\epsilon}{2}\nabla_1\log p_\phi (x^{k - 1}), \beta\epsilon\mathrm{Id}_n),\ k = 1,\dotsc,K,%
    \label{eq:langevin}
\end{equation}
where \( \mathcal{N}(\mu, \Sigma) \) denotes the normal distribution on \( \R^n \) with mean \( \mu \) and covariance \( \Sigma \). \( \beta, \epsilon \in \R^+ \) are appropriately chosen scaling parameters, and \( K \) denotes the total number of steps.
To aid the convergence of the Markov chains, we further follow the idea of persistent chains~\cite{tieleman_persistent_2008} and use a buffer in which the states of the chains persist throughout parameter updates.
\subsection{Experimental Setup}
For all the following experiments, we set \( n_f = \num{48} \), resulting in \( n_p = \num{12179905} \) and set the ReLU leak coefficient to \num{0.05}.
We trained the regularizer on the Low Dose CT Image and Projection dataset~\cite{moen_lowdose_2020}, where the images were downsampled to \( \num{128} \times \num{128} \).
We optimized~\eqref{eq:grad nll} using Adam~\cite{kingma_adam_2015} with a learning rate of \num{5e-4} and set the first and second order momentum variables to \( \beta_1 = \num{0.9} \) and \( \beta_2 = \num{0.999} \).
To stabilize training, we convolved \( p_x \) with \( \mathcal{N}(0, \sigma_{\text{data}}^2 \mathrm{Id}_n) \), where \( \sigma_{\text{data}} = \num{1.5e-2} \).
We used a batch size of \( \num{25} \) and a replay buffer holding \( \num{8000} \) images with reinitialization chance of \( p_{\text{re}} = \SI{1}{\percent} \).
Samples in the buffer were reinitialized with an equal chance of uniform noise or samples from the data distribution.
To sample \( p_\phi \), we ran~\eqref{eq:langevin} with \( K = \num{500} \), using \( \epsilon = 1 \) and \( \beta = \num{7.5e-3} \).\footnote{Similar to~\cite{nijkamp_shortrun_2019}, we reparametrize the regularizer as \( \frac{R}{T} \) for a-priori chosen \( T \), such that \( \epsilon = 1 \) in~\eqref{eq:langevin}.}
We summarize the training algorithm in Alg.~\ref{alg:training}.

For the reconstruction problems, we used accelerated proximal gradient descent~\cite{nesterov_agd}, as summarized in Alg.~\ref{alg:inference} with \( J = \num{1e3} \), \( \gamma_1 = 0.5 \), \( \gamma_2 = 1.5^{-1} \).
We solve the proximal operator \( \operatorname{prox} \colon \R^n \to \R^n \), which for \( H \colon \R^n \to \R \) and \( \tau \in \R^+ \) is defined as
\begin{equation}
	\operatorname{prox}_{\tau H} (y) = \argmin_{x} \tau H(x) + \frac{1}{2} \norm{x - y}_2^2
\end{equation}
using \( \num{10} \) iterations of the conjugate gradient method.
In what follows, the forward operator \( A \) assumes a parallel-beam geometry with \( n_d = \num{362} \) detectors of size \num{1} pixel and is discretized using the ASTRA toolbox~\cite{vanaarle_astra_2015}.
Unless stated otherwise, \( \eta \) is \( \SI{0.1}{\percent} \) Gaussian noise.
\begin{algorithm}[t]
	\DontPrintSemicolon%
	\SetKwInOut{Input}{Input}
	\SetKwInOut{Output}{Output}
	\Input{\( p_x \), \( \sigma_\text{data} \), \( n_\text{buffer} \), \( p_{\text{re}} \), \( K \), \( \phi \), \( n_e \), \( \epsilon \), \( \beta \)}
	\Output{\( \phi \) approximately minimizing \eqref{eq:nll}}
    \( \mathcal{B} \leftarrow \{ u_1,\dotsc,u_{n_\text{buffer}} \} \), \( u_i \sim \mathcal{U}({[0, 1]}^{n}) \)\;
	\For{\( t = 1,\dotsc,n_e\)}{
		\( x^+ \sim (p_x * \mathcal{N}(0, \sigma_{\text{data}}^2\mathrm{Id}_n)), x^0 \sim \mathcal{B} \)\;
		Generate \( x^- \) with~\eqref{eq:langevin} using \( x^0 \), \( \epsilon \), \( K \), \( \beta \)\;
		\lIf{%
            \( r > p_{\text{re}} \)
		}{%
		    \( x_{\text{refill}} = x^- \)
		}
		\Else{
			\lIf{%
			    \( r > 0.5 \)
		    }{%
		        \( x_{\text{refill}} = x^+ \)
		    }
		    \lElse{\( x_{\text{refill}} = u \sim \mathcal{U}({[0, 1]}^{n}) \)}
        }
		\( \mathcal{B} \leftarrow \mathcal{B} \setminus \{ x^0 \} \cup \{ x_{\text{refill}}\}  \)\;
		\( \phi \leftarrow \operatorname{Adam}(\nabla_{2} R(x^+, \phi) - \nabla_2 R(x^-, \phi)) \)\;
	}
	\caption{%
	    Maximum Likelihood training of an \gls{ebm}.
	    \( \mathcal{U}(\mathcal{X}) \) denotes the uniform distribution on \( \mathcal{X} \) and each \( r \) denotes an independent sample from \( \mathcal{U}([0, 1]) \).
    }%
	\label{alg:training}
\end{algorithm}
\begin{algorithm}[t]
	\DontPrintSemicolon%
	\SetKwInOut{Input}{Input}
	\SetKwInOut{Output}{Output}
	\SetKwRepeat{Do}{do}{while}
	\SetKw{Break}{break}
	\Input{initial \( \alpha \), \( f \), \( x^0 \), \( \phi \), \( J \), \( \gamma_1 \in (0, 1) \), \( \gamma_2 \in (0, 1) \)}
	\Output{\( x^{J + 1} \) approximately minimizing \eqref{eq:variational problem parametrized}}
	\( x^1 = x^0 \)\;
	\For{\( t = 1,\dotsc,J \)}{
		\( \bar{x} = x^{t} + \frac{t}{t + 3} (x^{t} - x^{t - 1}) \)\;
		\( g = \nabla_{1} R(\bar{x}, \phi) \)\;
		\For{ever}{%
		    \( x^{t + 1} = \operatorname{prox}_{\alpha D(\cdot, f)}(\bar{x} - \alpha g) \)\;
		    \( Q=R(\bar{x}, \phi) + \langle g, x^{t + 1} - \bar{x} \rangle + \frac{1}{2 \alpha} \norm{x^{t + 1} - \bar{x}}_2^2 \)\;
		    \uIf{\( R(x^{t+1}, \phi)  \leq Q \)}{\( \alpha \leftarrow \alpha / \gamma_1 \)\;\textbf{break}}
		    \lElse{\( \alpha \leftarrow \gamma_2\alpha \)}
		}
	}
	\caption{%
		Accelerated proximal gradient descent with Lipschitz-backtracking.
	}%
	\label{alg:inference}
\end{algorithm}

\section{RESULTS}
\subsection{Induced Prior Distribution}%
\label{ssec:prior}
For most hand-crafted regularizers, there typically exists a geometrical interpretation.
For instance, it is well known that \gls{tv} is related to the perimeter of the level sets of an image~\cite{ChambolleCasellesCremersNovagaPock+2010+263+340}.
Hence, the influence on the reconstruction is fairly easily understood.
Our regularizer can hardly be interpreted in such a way, however the energy-perspective allows for a statistical analysis by means of the Gibbs-Boltzmann distribution \( p_\phi \).

One of the main characteristics of any distribution are the points which locally maximize the density (modes).
By~\eqref{eq:gibbs} it is easily seen that the modes of \( p_\phi \) conincide with local minima of \( R(\cdot, \phi) \).
However, modes may occur as spikes in regions of generally low mass, and thus samples may represent the underlying distribution more accurately.
Therefore, we inspect our regularizer by computing modes as well as samples.

\begin{figure*}
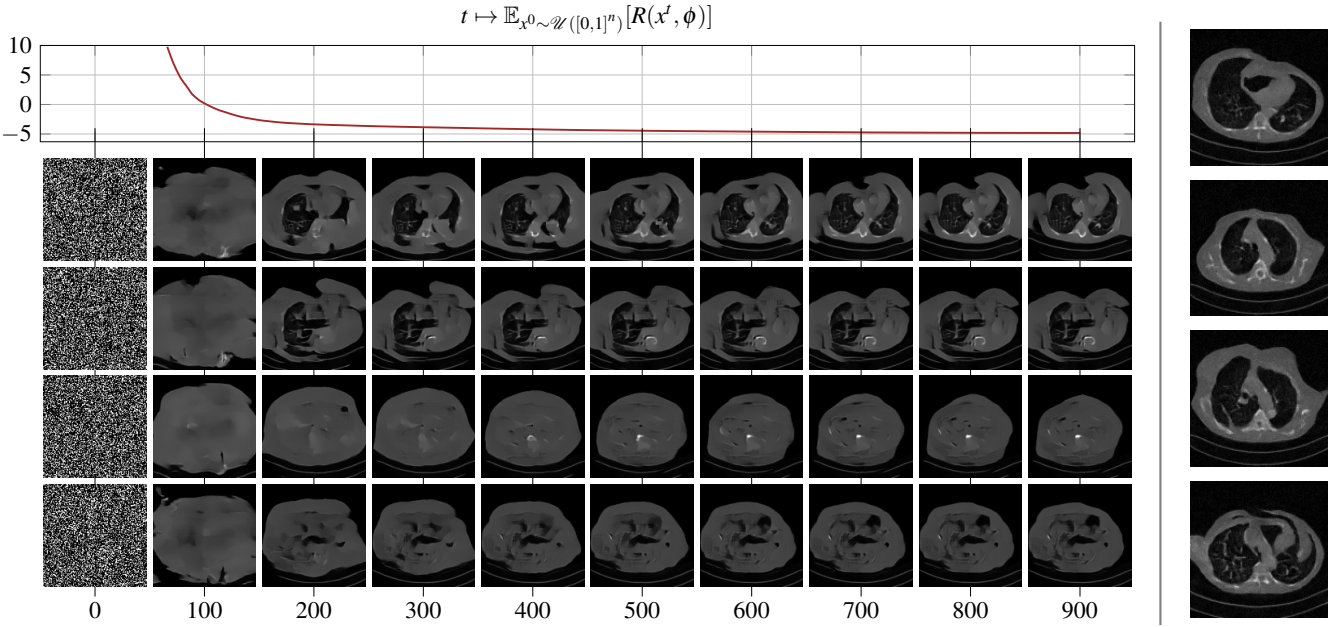

	\centering
	\scalebox{0.9}{\begin{tikzpicture}[trim axis left, trim axis right]
		\begin{axis}[
			height=3cm, width=17.58cm,
			restrict y to domain=-10:10, ymax=10,
			title={\( t \mapsto \mathbb{E}_{x^0\sim\mathcal{U}({[0, 1]}^n)}[R(x^{t}, \phi)] \)},
			grid=major,
			xmin=-50, xmax=950,
			xticklabels={},
			enlarge x limits=false, xtick={0,100,...,900},
			title style={yshift=-1ex}
		]
			\addplot [domain=0:900,mark=none,photoncolor,thick] table [x=idx, y=c, col sep=comma] {./figures/modes/energies.csv};
		\end{axis}
		\begin{scope}[xshift=-.8cm, yshift=-5.8cm]
			\foreach [count = \i] \step in {000, 100, 200, 300, 400, 500, 600, 700, 800, 900}
			{
				\pgfmathsetmacro{\xx}{\i * 1.6}
				\draw (\xx, 6) -- ++(0, -6.9);
				\node at (\xx, -1.1) {\num{\step}};
				\foreach \imagenum in {1, ..., 4}
				{
					\pgfmathsetmacro{\yy}{(\imagenum - 1) * 1.6}
					\pgfmathsetmacro{\xx}{\i * 1.6}
					\node at (\xx, \yy) {\includegraphics[width=1.5cm]{./figures/modes/\imagenum/\step.png}};
				}
			}
		\end{scope}
\end{tikzpicture}}
    \begin{tikzpicture}
        \node at (-1,0){};
        \draw[gray,thick] (-.9, 1) -- ++(0, -8);
        \foreach \i in {0, ..., 3}
        {
            \node at (.4, -2 * \i) {\includegraphics[width=1.8cm]{./figures/samples/\i.png}};
        }
    \end{tikzpicture}
	\caption{%
		Trajectories of the images from uniform noise to \( \argmin_x R(x, \phi) \) along with the corresponding \( R(x^{t}, \phi) \) (left) and samples \( x \sim p_\phi \) from the Langevin process~\eqref{eq:langevin} after \( K = \num{40000} \) steps (right).  
	}%
	\label{fig:trajectories}
\end{figure*}

We find \( x \sim p_\phi \) using Langevin sampling~\eqref{eq:langevin} with \( K = \num{40000} \) steps, and find \( \argmin_x R(x, \phi) \) with Alg.~\ref{alg:inference} using \( D(x,f) = 0 \).
In both cases, we set \( x_0 \sim \mathcal{U}({[0, 1]}^n) \).
We show the trajectories of \( x^t \) during minimzation of \( R(\cdot, \phi) \) and samples \( x \sim p_\phi \) in Fig. \ref{fig:trajectories}.

The results indicate that our model is able to synthesize natural \gls{ct} images without any measurement data.
This is in stark contrast to other priors typically used in medical imaging (see e.g.~\cite[Fig. 1]{adler_deep_2018} for samples drawn from hand-crafted priors).
\subsection{Limited-Angle and Few-View Reconstruction}
In this section, we shift our focus towards \gls{ct} reconstruction problems, where we first treat the reconstruction problem as a deterministic mapping in the \gls{map} sense.
Specifically, we denote by  \( \optimal{x} \colon \R^m \to \R^n \) the model-optimal reconstruction identified by the mapping
\begin{equation}
    \optimal{x}(f) \in \argmin_x \{ D(x, f) + R(x, \phi) \}.%
    \label{eq:variational map}
\end{equation}
Further, let \( p_{\hat{x}} \) denote a distribution on \( \R^n \times \R^m \) of (problem-dependent) \( (f, x) \)-pairs of a (noisy and incomplete) datum \( f \) and the corresponding reference image \( x \). 

\begin{figure}
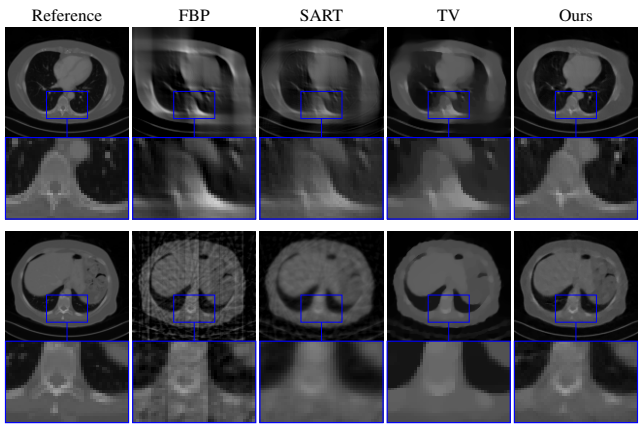

	\centering
	\scalebox{0.54}{
    \begin{tikzpicture}[font=\large]
		\foreach [count = \i] \method/\anno in {reference/Reference, fbp/{\gls{fbp}}, sart/{\gls{sart}}, tv/{\gls{tv}}, ours/{Ours}}
		{
            \begin{scope}[spy using outlines={rectangle, magnification=3, width=3cm, height=2cm, connect spies}]
                \pgfmathsetmacro{\xx}{\i * 3.1}
                \node at (\xx, 0) {\includegraphics[width=3cm]{./figures/map/limited-angle/\method.png}};
                \node at (\xx, -5) {\includegraphics[width=3cm]{./figures/map/few-view-20/\method.png}};
                \node at (\xx, 1.8) {\anno};
                \spy [blue] on (\i * 3.1, -.4) in node [left] at (\i * 3.1 + 1.5, -2.2);
                \spy [blue] on (\i * 3.1, -.4 - 5) in node [left] at (\i * 3.1 + 1.5, -2.2 - 5);
            \end{scope}
		}
	\end{tikzpicture}}
	\caption{%
		Comparison between \gls{fbp}, \gls{sart}, \gls{tv}, and our method for limited-angle (\( \theta \in [0, \frac{\pi}{2}] \), top) and few-view (\( n_\theta = 20 \), bottom) \gls{ct} reconstruction.
		Our model is able to faithfully reconstruct the image, whereas the other methods are not able to fully remove the smearing and streaking artifacts.
	}%
	\label{fig:map}
\end{figure}

To illustrate the capabilities of our trained regularizer, we first consider a limited-angle reconstruction problem.
Specifically, we reconstruct an image from \( n_\theta = \num{270} \) projections uniformly spaced over the quarter-circle \( \theta \in [0, \frac{\pi}{2}] \).
We show qualitative results in Fig.~\ref{fig:map} (top), where the \gls{fbp} reconstruction exhibits smearing artifacts that are characteristic of limited-angle \gls{ct}.
\gls{sart}~\cite{andersen_sart_1984} and additional \gls{tv} regularization help remedy this problem somewhat, however the reconstruction is not satisfactory.
We observe unnatural disconnected contours in the reconstruction, especially around the thorax.
On the contrary, our model is capable of reconstructing a natural looking image with realistic anatomy and high level of detail.
We show \( \mathbb{E}_{(f, x) \sim p_{\hat{x}}}[\operatorname{PSNR}(\optimal{x}(f), x)] \) in Tab.~\ref{tab:results}.
The results are in accordance with the qualitative analysis, with our model improving the \gls{tv} reconstruction by over \SI{4.5}{\decibel}.

In contrast to limited-angle \gls{ct}, in few-view \gls{ct} data are acquired over the full half-circle \( \theta \in [0, \pi] \).
However, on this half-circle only \( n_\theta \ll \) projections are sparsely acquired.
In traditional reconstruction algorithms, the sparse data manifests itself as streaking artifacts around sharp contours, where subsequent projections do not properly cancel each other.
Such artifacts can clearly be seen in the \gls{fbp} reconstruction in Fig.~\ref{fig:map} (bottom), where we show the results for a \( n_\theta = \num{20} \) few-view reconstruction problem.
\gls{tv} regularization yields a sharp and largely artifact-free image at the cost of losing almost all details.
Our method can reconstruct the image satisfactorily, where artifacts are removed whilst retaining small details.
Tab.~\ref{tab:results} shows quantitative results, with our approach consistently beating the reference methods for all \( n_\theta \in \{ \num{100}, \num{50}, \num{30}, \num{20} \} \) by a large margin.
\sisetup{detect-weight=true,detect-inline-weight=math}
\begin{table}
	\centering
	\caption{%
		\( \mathbb{E}_{(f, x) \sim p_{\hat{x}}}[\operatorname{PSNR}(\optimal{x}(f), x)] \) for limited-angle (\( \theta \in [0, \frac{\pi}{2}] \)) and few-view (\( n_\theta \in \{ 100, 50, 30, 20 \} \)) reconstruction.
	}%
	\label{tab:results}
	\begin{tabular}{lr*4S[table-format=2.2]}
		& & {\gls{fbp}} & {\gls{sart}} & {\gls{tv}} & {Ours}  \\\toprule
		limited-angle & \( \theta \in [0, \frac{\pi}{2}] \) & 19.05    & 27.72     & 29.67     & \bfseries34.21 \\\midrule
		\multirow{4}{*}{few-view}& \( n_\theta=100 \) &  37.15   &   43.86   &   46.77   & \bfseries 49.47 \\
		& \( n_\theta=50 \) &  33.12    &   37.05   &   40.21   & \bfseries 45.06 \\
		& \( n_\theta=30 \) &  28.78    &   33.04   &   35.33   & \bfseries 41.65 \\
		& \( n_\theta=20 \) &  25.24    &   30.55   &   31.77   & \bfseries 38.48 \\\bottomrule
	\end{tabular}
\end{table}
\subsection{Posterior Analysis}
\begin{figure}
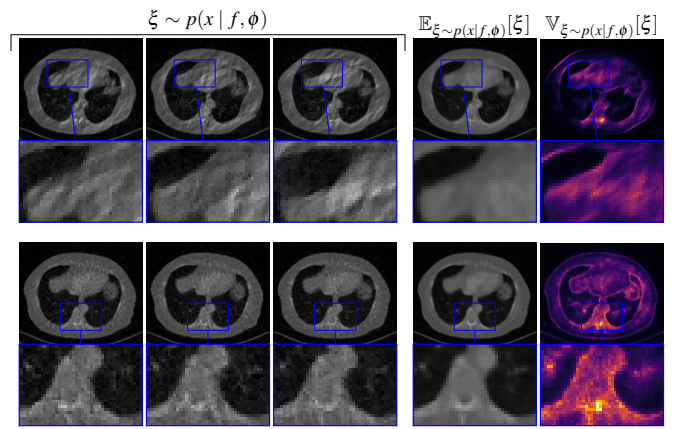

	\centering
	\scalebox{0.54}{\begin{tikzpicture}[font=\Large]
		\def\imwidth{3}
		\def\ppad{0.1}
		\foreach \problem/\yoff/\xxx/\yyy in {limited-angle/0cm/0.4/0.7, few-view-30/-5cm/0.5/0.4}
		{
			\begin{scope}[yshift=\yoff,]
			    \foreach [count = \timecount] \ttime in {100, 150, 199} {
			        \begin{scope}[spy using outlines={rectangle, lens={scale=3}, width=3cm, height=2cm, connect spies}]
                        \pgfmathsetmacro{\xx}{\timecount * (\imwidth + \ppad)}
                        \node (im \problem \timecount) at (\xx, 0) {\includegraphics[width=\imwidth cm]{./figures/posterior-sampling/\problem/\ttime.png}};
                        \spy [blue] on (image cs:image=im \problem \timecount, x=\xxx, y=\yyy) in node [left] at (\timecount * \imwidth + \timecount * \ppad + 1.5, -2);
                    \end{scope}
                }
                \begin{scope}[spy using outlines={rectangle, lens={scale=3}, width=3cm, height=2cm, connect spies}]
                    \pgfmathsetmacro{\xx}{4 * (\imwidth + \ppad) + 0.3}
                    \node (im \problem mean) at (\xx, 0) {\includegraphics[width=\imwidth cm]{./figures/posterior-sampling/\problem/mean.png}};
                    \spy [blue] on (image cs:image=im \problem mean, x=\xxx, y=\yyy) in node [left] at (4 * \imwidth + 4 * \ppad + 1.5+0.3, -2);
                    \pgfmathsetmacro{\xx}{5 * (\imwidth + \ppad) + 0.3}
                    \node (im \problem std) at (\xx, 0) {\includegraphics[width=\imwidth cm]{./figures/posterior-sampling/\problem/std.png}};
                    \spy [blue] on (image cs:image=im \problem std, x=\xxx, y=\yyy) in node [left] at (5 * \imwidth + 5 * \ppad + 1.5+0.3, -2);
                \end{scope}
            \end{scope}
		}
		\pgfmathsetmacro{\xxleft}{1 * (\imwidth + \ppad) - .2 - \imwidth/2}
		\pgfmathsetmacro{\xxright}{3 * (\imwidth + \ppad) + .2 + \imwidth/2}
		\draw (\xxleft, 1.2) -- ++(0, 0.4) -- node [midway, above] {\(\xi \sim p(x \mid f, \phi) \)} (\xxright, 1.6) -- ++(0, -0.4);
        \pgfmathsetmacro{\xx}{4 * (\imwidth + \ppad) + 0.3}
		\node at (\xx, 1.8) {\( \mathbb{E}_{\xi \sim p(x \mid f, \phi)}[\xi] \)};
        \pgfmathsetmacro{\xx}{5 * (\imwidth + \ppad) + 0.3}
		\node at (\xx, 1.8) {\( \mathbb{V}_{\xi \sim p(x \mid f, \phi)}[\xi] \)};
	\end{tikzpicture}}
	\caption{%
		Sampling the posterior of a limited-angle (\( \theta \in [0, \frac{\pi}{2}]\), top) and few-view (\( n_\theta = 30 \), bottom) \gls{ct} reconstruction problem:
		The three images on the left show different samples during the sampling process, the two images on the right show the expected value and variance of the posterior distribution respectively.
	}%
	\label{fig:posterior sampling}
\end{figure}

Instead of treating \( R \) as a point estimator in the maximum a-posteriori sense~\eqref{eq:variational map}, the Bayesian formulation allows to explore the full posterior distribution of any given reconstruction problem.
This is especially useful in the medical domain, where interpretability is of utmost importance.
To this end, we perform Langevin sampling of the posterior distribution~\eqref{eq:bayes} with the same parameters as in training.
We show some illustrative examples for limited-angle and few-view \gls{ct} in Fig.~\ref{fig:posterior sampling}.
The figure shows samples \( \xi \sim p(x \mid f, \phi)=p_\phi(x)p(f \mid x) \) from the posterior distribution associated with Eq.~\eqref{eq:variational problem parametrized} as well as it's expectation and variance.
For the limited-angle reconstruction, we observe large variance around regions of high ambiguity, where there exist no projections to define contours.
Similarly, for the few-view problem, there is high variance around small structures such as the vertebrae or blood vessels in the lung.
For both problems, the approximated expected value over the posterior also yields a visually appealing, although somewhat over-smoothed, reconstruction.
\subsection{Out-of-Distribution Application}
\subsubsection{Uncertainty Quantification Through Posterior Variance Analysis}
To study how the variance relates to uncertainty, we perform the following experiment:
We introduce unnatural (read: not present in the training data) structures into the image by overlaying the \enquote{cameraman} image and an example of the \enquote{grid} texture from the Describable Textures Dataset~\cite{cimpoi14describing} on a reference scan.
Subsequently, we approximate the variance of a few-view reconstruction problem using \( n_\theta = \num{20} \) views by Langevin sampling.

\begin{figure}
	\centering
    \begin{tikzpicture}[font=\footnotesize]
		\def\imwidth{2.7}
		\def\ppad{0.1}
		\foreach [count=\probcount] \problem/\yoff in {cameraman/0cm, grid/-2.75cm, clean/-5.8cm}
		{
			\begin{scope}[yshift=\yoff,]
			    \foreach [count = \timecount] \ttime/\annotation in {scan/Reference, mean/\( \mathbb{E}_{\xi \sim p(x \mid f, \phi)}[\xi] \), std/\( \mathbb{V}_{\xi \sim p(x \mid f, \phi)}[\xi] \)}
			    {
			        \pgfmathsetmacro{\xx}{\timecount * (\imwidth + \ppad)}
                    \node [inner sep=0,outer sep=0] at (\xx, 0) {\includegraphics[width=\imwidth cm]{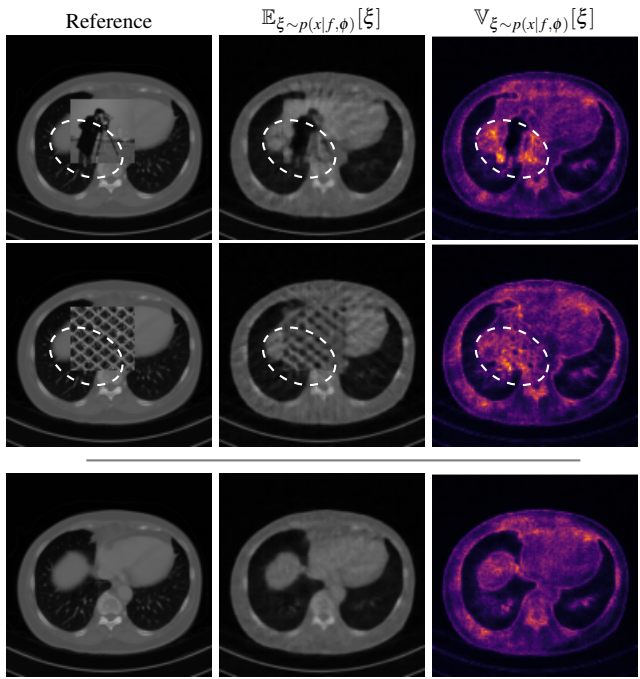}};
                    \ifthenelse{\probcount=1}{
                        \node at (\xx, 1.55) {\annotation};
                    }{}
			        \ifthenelse{\probcount<3}{
                        \node [draw, dashed, thick, white, ellipse, minimum width=1cm, minimum height=0.7cm, rotate=-25] at (\xx - 0.3, -.15) {};
                    }{}
                }
            \end{scope}
		}
		\draw [gray, thick] (2.5, -4.28) -- ++(6.5, 0);
	\end{tikzpicture}
	\caption{%
        Comparison of the posterior distribution of a corrupted (top two) versus clean (bottom) scan:
        The high variance around the corrupted regions (highlighted) relates to high model uncertainty.
	}%
	\label{fig:posterior comparison}
\end{figure}

We show the expected value and variance over the posterior for the clean and corrupted scans in Fig.~\ref{fig:posterior comparison}.
Although the bulk of the cameraman shows low variance (and indeed the reconstruction looks natural in these regions), we observe high variance in unnatural regions, such as the artificially introduced corners and the tripod.
Similarly, compared to the reference scan the grid overlay leads to high variance in the posterior. 

In general, we believe that high posterior variance is related to model uncertainty.
To be more specific, we expect high variance if the measurement data suggests structures that are not consistent with the training data.
This could potentially aid in detecting pathologies in images.
\subsubsection{Generalization}
\begin{figure}
	\centering
    \begin{tikzpicture} [font=\footnotesize,>=latex]
		\begin{axis}[
			ymin=27,
			ymax=34,
			grid=major,
			width=\textwidth,
			height=4cm,
			width=9cm,
			title={\( \kappa \mapsto \mathbb{E}_{x \sim p_x} [\operatorname{PSNR}(\optimal{x}(x_\kappa), \operatorname{rot}_\kappa (x))] \)},
			name=plot,
			title style={yshift=-1ex},
		]
			\addplot [photoncolor, mark=x] table [x=x,y=y] {
				x    y
				0    33.39
				1    32.62
				2    31.81
				3    31.43
				4    30.73
				5    30.25
				10   29.62
				15   28.50
				20   28.33
				25   28.17
				30   28.09
				35   28.35
				40   28.20
			};
			\node (kappa0) [draw, rectangle, light] at (axis cs:0,33.39) {};
			\node (kappa5) [draw, rectangle, light] at (axis cs:5,30.25) {};
			\node (kappa40) [draw, rectangle, light] at (axis cs:40,28.20) {};
		\end{axis}
		\begin{scope}[xshift=1.5cm, yshift=-1.8cm]
			\foreach \offset/\kkappa in {0/0, 2.8/5, 5.6/40}
			{
				\node (\kkappa) [light, draw, rectangle, minimum width=2.5cm, minimum height=2.5cm] at (\offset-0.6, 0) {\includegraphics[width=2.3cm]{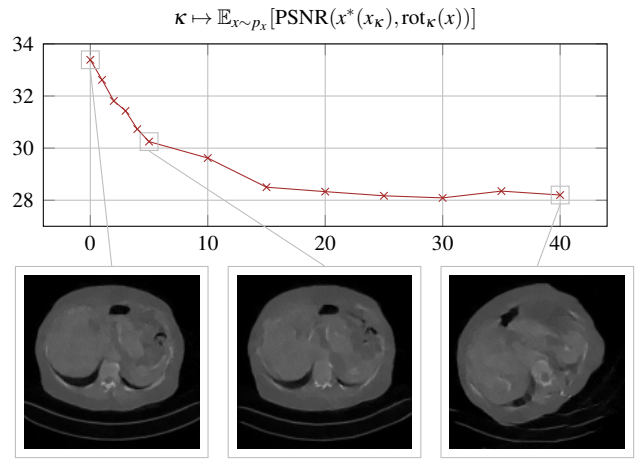}};
				\draw [light] (kappa\kkappa.south) -- (\kkappa.north);
			}
		\end{scope}
	\end{tikzpicture}
	\caption{%
		Performance of the regularizer on out-of-distribution data:
		For denoising rotated images, the \gls{psnr} quickly decays even for small rotations.
	}%
	\label{fig:ood-plot}
\end{figure}

In Sec.~\ref{ssec:prior} we have shown how samples \( x \sim p_\phi \) resemble data drawn from \( p_x \) --- that is, \( R \) encodes a prior in the frequentist sense.
With this, a natural question is if our proposed regularizer can be applied to reconstruction problems where the underlying distribution deviates far from \( p_x \).
To study this, we propose the following experiment:
We let
\begin{equation}
    x_{\kappa} = \operatorname{rot}_{\kappa} (x) + \eta,
\end{equation}
where \( \operatorname{rot}_{\kappa} \colon \R^n \to \R^n \) is the bi-linear rotation operator of angle \( \kappa \) and \( \eta \) is \SI{10}{\percent} Gaussian noise and find
\begin{equation}
    \optimal{x} \in \argmin_{x} \frac{1}{2\sigma^2}\norm{x - x_{\kappa}}^2 + R(x, \phi).
\end{equation}
The results in Fig.~\ref{fig:ood-plot} show that performance quickly deteriorates with increasing \( \kappa \).
This is in line with our expectations, since our regularizer models global characteristics of the reconstruction which are not rotation invariant.
\section{CONCLUSION}
In this work, we designed a parametrized regularizer utilizing a global receptive field, which we trained on full-scale \gls{ct} images by maximizing their likelihood.
The induced Gibbs-Boltzmann distribution of the trained regularizer strongly resembles the data distribution --- that is, our model is capable of synthesizing natural \gls{ct} images \emph{without} any data.
The maximum likelihood framework does not assume any particular forward acquisition operator or noise statistics, and the trained regularizer can be applied to any reconstruction problem.
In limited-angle and few-view reconstruction problems, we observed significantly improved quantitative and qualitative performance when compared to classical reconstruction algorithms.
Further, we were able to relate the variance in the posterior with unnatural structures in the underlying image, as is the case for certain pathologies.

In summary, we believe that learning energy-based models capable of truly capturing the underlying distribution is a very promising direction for future research.
Such models yield natural reconstructions with severely undersampled and noisy data, where data consistency can be enforced with arbitrary data terms.
We also want to emphasize that training requires only reconstructed images, which are typically much more ubiquitous than image-data pairs.
Future work includes the extension to higher resolutions used in clinical practice today, and tackling the problem of scale- and rotation-invariance.
Further, a rigorous mathematical analysis in the context of inverse problems, stability w.r.t.\ training and measurement data would improve the applicability in clinical practice.

{\small\bibliographystyle{IEEEtranS}\bibliography{bibliography}}
\end{document}

%% file: acronyms.tex
\newacronym{ai}{{AI}}{Artificial Intelligence}
\newacronym{art}{{ART}}{Algebraic Reconstruction Technique}
\newacronym{sirt}{{SIRT}}{Simultaneous Iterative Reconstruction Technique}
\newacronym{sart}{{SART}}{Simultaneous Algebraic Reconstruction Technique}
\newacronym{bicav}{{BICAV}}{Block Iterative Component Averaging}
\newacronym{cgls}{{CGLS}}{Conjugate Gradient Least Squares}
\newacronym{ossqs}{{OS-SQS}}{Ordered Subset Separable Quadratic Surrogates}
\newacronym{ann}{{ANN}}{Artificial Neural Network}
\newacronym{bp}{{BP}}{Backprojection}
\newacronym{admm}{{ADMM}}{Alternating Direction Method of Multipliers}
\newacronym{cnn}{{CNN}}{convolutional neural network}
\newacronym{cg}{{CG}}{Conjugate Gradient}
\newacronym{cs}{{CS}}{Compressed Sensing}
\newacronym{ct}{{CT}}{Computed Tomography}
\newacronym{cpu}{{CPU}}{Central Processing Unit}
\newacronym{dicom}{{DICOM}}{Digital Imaging and Communications in Medicine}
\newacronym{dof}{{DOF}}{Degrees of Freedom}
\newacronym{ddr}{{DDR}}{Digitally Reconstructed Radiograph}
\newacronym{dsc}{{DSC}}{Dice Similarity Coefficient}
\newacronym{dti}{{DTI}}{Diffusion Tensor Imaging}
\newacronym{ecg}{{ECG}}{Electrocardiography}
\newacronym{em}{{EM}}{Expectation Maximization}
\newacronym{ft}{{FT}}{Fourier Transform}
\newacronym{fft}{{FFT}}{Fast Fourier Transform}
\newacronym{fista}{{FISTA}}{Fast Iterative Shrinkage and Thresholding Algorithm}
\newacronym{fbp}{{FBP}}{Filtered Back-Projection}
\newacronym{foe}{{FoE}}{Fields of Experts}
\newacronym{fov}{{FoV}}{Field of View}
\newacronym{gac}{{GAC}}{Geodesic Active Contours}
\newacronym{gan}{{GAN}}{Generative Adversarial Network}
\newacronym{gd}{{GD}}{Gradient Descent}
\newacronym{gmm}{{GMM}}{Gaussian Mixture Model}
\newacronym{gpu}{{GPU}}{Graphics Processing Unit}
\newacronym{hu}{{HU}}{Hounsfield Units}
\newacronym{ista}{{ISTA}}{Iterative Shrinkage and Thresholding Algorithm}
\newacronym{iipg}{{IIPG}}{Inertial Incremental Proximal Gradient}
\newacronym{ipalm}{{IPALM}}{Inertial Proximal Alternating Linearized Minimization}
\newacronym{lsc}{{l.s.c.}}{lower-semicontinuous}
\newacronym{lista}{{LISTA}}{Learned Iterative Shrinkage and Thresholding Algorithm}
\newacronym{lbfgs}{{L-BFGS}}{Limited-Memory Broyden-Fletcher-Goldfarb-Shanno}
\newacronym{map}{{MAP}}{maximum a-posteriori}
\newacronym{mlp}{{MLP}}{Multi Layer Perceptron}
\newacronym{mr}{{MR}}{Magnetic Resonance}
\newacronym{ml}{{ML}}{Maximum Likelihood}
\newacronym{mri}{{MRI}}{Magnetic Resonance Imaging}
\newacronym{mae}{{MAE}}{Mean Absolute Error}
\newacronym{mse}{{MSE}}{Mean Squared Error}
\newacronym{msssim}{{MS-SSIM}}{Multi-Scale Structural Similarity Index}
\newacronym{ncc}{{NCC}}{Normalized Cross Correlation}
\newacronym{nlm}{{NLM}}{Non-Local Means}
\newacronym{nufft}{{NUFFT}}{Non-Uniform Fast Fourier Transform}
\newacronym{nrmse}{{NRMSE}}{Normalized Root Mean Squared Error}
\newacronym{icp}{{ICP}}{Iterative Closest Point}
\newacronym{pat}{{PAT}}{Photoacoustic Tomography}
\newacronym{pca}{{PCA}}{Principal Component Analysis}
\newacronym{pet}{{PET}}{Positron Emission Tomography}
\newacronym{psf}{{PSF}}{Point Spread Function}
\newacronym{psnr}{{PSNR}}{Peak Signal-To-Noise Ratio}
\newacronym{rbf}{{RBF}}{Gaussian radial basis function}
\newacronym{relu}{{ReLU}}{Rectified Linear Unit}
\newacronym{roi}{{ROI}}{Region Of Interest}
\newacronym{snr}{{SNR}}{Signal-to-Noise Ratio}
\newacronym{sota}{SotA}{State-of-the-Art}
\newacronym{spect}{{SPECT}}{Single Photon Emission Computed Tomography}
\newacronym{ssim}{{SSIM}}{Structural Similarity Index}
\newacronym{tof}{{ToF}}{Time of Flight}
\newacronym{tgv}{{TGV}}{Total Generalized Variation}
\newacronym{tv}{{TV}}{Total Variation}
\newacronym{us}{{US}}{Ultrasound}
\newacronym{vn}{{VN}}{Variational Network}
\newacronym{sbp}{{SBP}}{Simple Back-Projection}
\newacronym{fdk}{{FDK}}{Feldkamp-Davis-Kress}
\newacronym{aec}{{AEC}}{Automatic Exposure Control}
\newacronym{bm3d}{{BM3D}}{Block Matching and 3D Filtering}
\newacronym{mcmc}{{MCMC}}{Markov Chain Monte Carlo}
\newacronym{cd}{{CD}}{Contrastive Divergence}
\newacronym{poe}{{PoE}}{Products of Experts}
\newacronym{mala}{{MALA}}{Metropolis adjusted Langevin algorithm}
\newacronym{lmc}{{LMC}}{Langevin Monte Carlo}
\newacronym{ula}{{ULA}}{unadjusted Langevin algorithm}
\newacronym{tdv}{{TDV}}{Total Deep Variation}
\newacronym{ebm}{{EBM}}{Energy-based model}